\documentclass[epjST]{svjour}
\usepackage{graphics,epsfig}
 \begin{document}
 \title{The spin-flip amplitude \\ in the impact-parameter representation}
 \author{O.\,V. Selyugin\inst{1}\fnmsep\thanks{\email{selugin@theor.jinr.ru}}
 \and J.-R. Cudell\inst{2}\fnmsep\thanks{\email{JR.Cudell@ulg.ac.be}}
}
 \institute{JINR, Bogoliubov Laboratory of Theoretical Physics,
 141980~Dubna, Moscow Region, Russia
 \and IFPA, AGO Dept., Universit\'e de Li\`{e}ge , Belgium }
 \abstract{
   The impact-parameter representation of the spin-flip amplitude
   of had\-ron elastic scattering is examined in different unitarisation
   schemes, taking  the Born term of the spin-flip amplitude
   from the  Dubna Dynamical Model (DDM).
 It is shown that the basic properties of the unitarisation schemes
  are independent of the functional form used for unitarisation
  but heavily depend on the asymptotic value of the unitarised amplitude.
} 

\maketitle
 \section{Introduction}

 There are many different models for the description of hadron elastic
 scattering at small angles \cite{mog1,selmog1}. They lead to  different
 predictions for the structure of the scattering amplitude at asymptotic
 energies, where the diffraction  processes can display complicated
 features \cite{dif04}.  This concerns especially the asymptotic unitarity
 bound connected with the Black Disk Limit (BDL). In this paper, we study
 the impact of unitarisation on the spin properties of the elastic amplitude.

 In practice,  we need to sum many different waves with
 $l \rightarrow \infty $ and this leads to the impact parameter
 representation \cite{Predazzi66} converting
 the summation over $l$  into an integration over the impact parameter $b$.
 In the  impact-parameter representation, the  Born term of the scattering amplitude will be
 \begin{eqnarray}
 \chi(s,b) \  \approx 
   \ \int \ d^2 q \ e^{i \vec{b} \vec{q} } \  F_{\rm Born}\left(s,q^2\right)\,,
 \label{tot02}
 \end{eqnarray}
 where we have dropped the kinematical factor $1/\sqrt{s(s-2m_p^2)}$
 and a factor $s$ in front of the scattering amplitude.

 The hadron spin-flip amplitude is expected to be
 connected with quark exchange between the scattering hadrons,
 and at high energy and small angles it is expected to be negligible.
 However, some models, which take into account non-perturbative
 effects, lead to a non-vanishing hadron spin-flip amplitude \cite{mog2,DDM-spin}
 even at high energy.

  After  unitarisation, we get for the scattering amplitude
  \begin{eqnarray}
 F(s,t)  \approx 
    \ \int \ e^{i \vec{b} \vec{q} }  \ \Gamma(s,b)   \ d^2 b\,,
 \label{overlap}
 \end{eqnarray}
  where $t= -q^2$.
 The overlap function $\Gamma(s,b)$ can be a matrix,
 corresponding to the scattering of different spin states.
 Unitarity of the $S$-matrix, $SS^{+}=1$, requires that $\Gamma(s,b) \leq 1$.

 The unitarisation procedure can be obtained in different
 ways, starting from simple diagrams in the  tree approximation and using
 the Shr\"odinger equation \cite{Shiff,Goldberger} or including the spin of
 particles in the relativistic case \cite{Blokhintzev,Barbashev}.
 One must take into account many additional diagrams which include
 inelastic states in the $s$ channel. This leads to different
 schemes, see for example
 \cite{Martiros,Kaidalov1,Martynov1} in which renormalised eikonal
 representations were obtained. No one really knows which are the
 leading diagrams and how to sum them. Therefore, all these
 approaches are possible phenomenological forms which can lead
 to different spin correlations.

 There are two  important constraints which any unitarisation scheme
 must satisfy.
 Firstly, in the limit of small energies, every unitarisation
 representation  must reduce to the same
 scattering amplitude $-$ the Born term $-$
 and thus must give the same result.
      Only at high energies, when the number of diagrams and their forms
 are essentially
     different, will the various unitarisation representations give
 different results \cite{CS-sat06,CS-BDL06}.

 Secondly, for any unitarisation scheme, the corresponding overlap
 function cannot exceed the unitarity bound. In different normalisations,
 this bound may  equal to $1$ or $2$.

 At LHC energies, the effects of unitarisation will be large and
 the experimental data will probably determine what form of
 unitarisation is realised.

 \section{Spin-dependent  scattering amplitude }

 In the case of elastic scattering of a baryon of momentum $p_1$
 on another baryon of momentum $k_1$, going to states of respective
 momenta $p_2$ and $k_2$, {\it e.g.}
 $pp \rightarrow pp,\  \bar{p}p \rightarrow \bar{p}p,\ np \rightarrow np,
 \ p\Lambda \rightarrow p\Lambda,\ \Lambda \Sigma \rightarrow \Lambda \Sigma,
 $
 the full representation for the  scattering amplitude is
 \begin{eqnarray}
 \Phi(s,t) &  =&
 \Phi_{1}(s,t)\bar{u}(p_2)u(p_1)\bar{u}(k_2)u(k_1) +
 \Phi_{2}(s,t)\bar{u}(p_2)\gamma K u(p_1)\bar{u}(k_2)\gamma P u(k_1) \nonumber \\
 &+&
 \Phi_{3}(s,t)\bar{u}(p_2)\gamma_{5} (\gamma K)u(p_1)\bar{u}(k_2)\gamma_{5}(\gamma P) u(k_1) +
 \Phi_{4}(s,t)\bar{u}(p_2)\gamma_{5} u(p_1)\bar{u}(k_2)\gamma_{5} u(k_1)  \nonumber \\
 &+&
 \Phi_{5}(s,t)\left[\bar{u}(p_2)\gamma K u(p_1)\bar{u}(k_2)u(k_1) +
 \bar{u}(p_2)u(p_1)\gamma P u(k_1)\bar{u}(k_2)\right] \nonumber \\
 &+&
 \Phi_{6}(s,t)\left[\bar{u}(p_2)\gamma K u(p_1)\bar{u}(k_2)u(k_1) -
 \bar{u}(p_2)u(p_1)\gamma P u(k_1)\bar{u}(k_2)\right] \nonumber \\
 &+&
 \Phi_{7}(s,t)\bar{u}(p_2)\gamma_{5} u(p_1)\bar{u}(k_2)\gamma_{5}(\gamma P) u(k_1)    \nonumber \\
 &+&    \Phi_{8}(s,t)\left[\bar{u}(p_2)\gamma_{5} (\gamma K) u(p_1)
 \bar{u}(k_2)\gamma_{5} u(k_1)\right]\,.
 \label{eik2}
 \end{eqnarray}
 Here
 \begin{eqnarray}
    K=\frac{1}{2}(k_1+k_2)\,; \ \ \ P=\frac{1}{2}(p_1+p_2)\,; \ \ \ Q=(k_2-k_1)=(p_2-p_1)\,.
 \end{eqnarray}
 The last two terms of (\ref{eik2})
 do not satisfy charge and time invariance and must be zero.
 If all four particles are  identical, the amplitude does not change under
 the exchange $p_1, p_2\rightleftharpoons k_1, k_2$ and  $P
 \rightleftharpoons K$. Hence, for  proton-proton scattering
 there are only five helicity amplitudes \cite{Lehar}:
  \begin{eqnarray}
    \Phi^{B}_{1} (s,t) &=& <++|++>\,; \ \ \  \Phi^{B}_{2} (s,t) = <++|-->\,; \nonumber \\
   \Phi^{B}_{3} (s,t) &=& <+-|+->\,; \ \ \  \Phi^{B}_{4} (s,t) = <+-|-+>\,; \ \ \
   \Phi^{B}_{5} (s,t) = <++|+->\,.
 \end{eqnarray}

 In the  Regge limit, $t$ fixed and $s \rightarrow \infty$,
 one can write the Regge-pole contributions to  the helicity amplitudes in
 the $s$-channel as \cite{Capella}
 \begin{eqnarray}
    \Phi^{B}_{\lambda_1, \lambda_2,\lambda_3, \lambda_{4}} (s,t) \approx
    \sum_{i} g^{i}_{\lambda_1, \lambda_2}(t)  g^{i}_{\lambda_3, \lambda_4}(t)
    [\sqrt{|t|}]^{|\lambda_1- \lambda_2|+|\lambda_3- \lambda_4|}
    \left(\frac{s}{s_0}\right)^{\alpha_{i}} (1 \pm e^{-i \pi \alpha_{i} })\,.
 \end{eqnarray}

 The differential cross sections is then given by
 \begin{eqnarray}
   \frac{d \sigma}{dt} = \frac{2 \pi}{s^2} \left( |\Phi_{1}|^2 + |\Phi_{2}|^2+
   |\Phi_{3}|^2 + |\Phi_{4}|^2+4 |\Phi_{5}|^2\right)\,.
 \end{eqnarray}
  The total helicity amplitudes can be written as $\Phi_{i}(s,t) =
  \Phi^{h}_{i}(s,t)+\Phi^{\rm em}_{i}(s,t) e^{\varphi(s,t)}$\,, where
 $\Phi^{h}_{i}(s,t)$ comes from the strong interactions,
 $\Phi^{\rm em}_{i}(s,t)$ from the electromagnetic interactions and
 $\varphi(s,t)$
 is the interference phase factor between the electromagnetic and strong
 interactions \cite{WY,Selphase}.

 The spin correlation parameters, the analysing power - $A_N$ and the
  and double-spin parameter $A_{NN}$, can be extracted from experimental measurements:
 \begin{eqnarray}
  A_{N} &=&\frac{ \sigma(\uparrow)- \sigma(\downarrow)}{\sigma(\uparrow)+ \sigma(\downarrow)} =
 \frac{\Delta \sigma^{s}}{\sigma_{0}}\,,\\
  A_{NN} &=&\frac{ \sigma(\uparrow \uparrow) - \sigma(\uparrow \downarrow)}{\sigma(\uparrow \uparrow)+
 \sigma(\uparrow \downarrow) } = \frac{\Delta \sigma^{d}}{\sigma_{0}}\,,
 \end{eqnarray}
 where $\Delta \sigma^{s}$ and $\Delta \sigma^{d}$ refer to the difference of
 single- and double-spin-flip cross sections.
 The expressions for these parameters will be
 \begin{eqnarray}
  A_{N} \frac{d \sigma}{dt} &=& -\frac{4 \pi}{s^2}
   \left [ Im (\Phi_{1} + \Phi_{2}+ \Phi_{3} - \Phi_{4}) \Phi^{*}_{5}\right ]\,;\\
  A_{NN} \frac{d \sigma}{dt} &=& \frac{4 \pi}{s^2}
  \left [ Re (\Phi_{1} \Phi^{*}_{2} - \Phi_{3} \Phi^{*}_{4}) + |\Phi_{5}|^2\right ]\,.
 \end{eqnarray}

 Regge factorisation together with the experimental information
 about the spin-correlation effects at  high energy and small momentum transfer
 in proton-proton  elastic scattering, suggest that the double
 helicity flip is a second-order effect and consequently that
 one can neglect the amplitudes $\Phi^B_{++--} (s,t)$ and  $\Phi^B_{+--+} (s,t)$.
 Furthermore, when the exchanged Regge trajectories have natural parity, we have
 for spin-non-flip amplitudes \cite{Morel}
  $\Phi^B_{++++} (s,t) = \Phi^B_{+-+-} (s,t)$.

 \section{\boldmath $U_e$-matrix unitarisation}
  The form  of  the  unitarisation of the  scattering amplitude
  in the impact parameter representation depends on the
 non-linear processes which lead to the saturation of the gluon density.
 There are many approaches to the equations describing such processes \cite{mog3}.
  The most popular one, the  dipole-dipole interaction model \cite{d-d-m}, describes the
   process of saturation as a function of the dipole size.
   Its inclusion into a real hadron-hadron interaction requires a phenomenological model.

   Here we shall consider only non-linear equations
    which  lead to the known unitarisation schemes.
   One of the simplest such equation is the well-known  logistic equation \cite{dif04,CPS-Prag},
 used for long time in many different branches of physics:
 \begin{eqnarray}
 \frac{dN}{dy} =\Delta N \ [ \ 1 - N \ ]\,, \label{log-eq}
 \end{eqnarray}
  where $y =\log(s/s_0)$ and $\Delta = 1- \alpha(0)$, ($\alpha(0)$
 being the intercept of the leading pole).

   Its solution has the form
 \begin{eqnarray}
   N = \frac{\chi(s,b)}{1 \ + \ \chi(s,b)}\,,
 \end{eqnarray}
 where $\chi(s,b)\approx s^{\Delta}$ is connected with the Born term of the scattering amplitude.

 Then  the scattering amplitude is
 \begin{eqnarray}
 \Phi^{h}(s,t) \ = \  \frac{i}{2  \pi}
   \ \int \ d^2 b  \  e^{i \vec{b} \vec{q} } \frac{\chi(s,b)}{1+ \chi(s,b)}\,.
 \label{K-mat}
 \end{eqnarray}
 This unitarisation scheme gives results similar to those of the
 eikonal representation, and we will refer to it as
 $U_e$-unitarisation.

 The  phase $\chi(s,b)$  is connected to  the interaction
 quasi-potential which can have real and imaginary parts and,
 in the case of a spin-dependent potential,
 a matrix structure:
 \begin{eqnarray}
 \chi(s,b)\ = \ F_{\rm Born}(s,b)  \ \approx   \frac{1}{k}
   \ \int  \hat{V}\left( \sqrt{b^2  + z^2}  \right) dz.
 \label{potential3}
 \end{eqnarray}
 If the quasi-potential contains a non-spin-flip part and, for example,
 spin-orbital and spin-spin interactions, the phase will be
 \begin{eqnarray}
  \chi(s,b) = \chi_{0}(s,b)
 - i \ \vec{n} \cdot (\vec{\sigma}_{1} + \vec{\sigma}_{2} ) \chi_{\rm LS}(s,b)
 - i  (\vec{\sigma}_{1} \cdot \vec{\sigma}_{2} ) \ \chi_{\rm SS}(s,b).
 \end{eqnarray}

 If we take into account only the spin-flip and spin-non-flip parts
 and neglect the second order on the spin-flip amplitude, the overlap function will be
 \begin{eqnarray}
 \Gamma(s,b)=\frac{\chi_{0}(s,b) +\sigma \chi_{sf}(s,b)}{1
 +\chi_{0}(s,b) +\sigma \chi_{sf}(s,b)}= 1- \frac{\left(1+\chi_{0}(s,b)\right) -
 \sigma \chi_{sf}(s,b) }{\left(1 + \chi_{0}(s,b)\right)^2 - \left(\sigma \chi_{sf}(s,b)\right)^2}\,.
 \label{K-m-s}
 \end{eqnarray}

 Using the representation for the Bessel functions
 \begin{eqnarray}
 J_{0}(x) \ =  \frac{1}{2 \pi}
   \ \int_{0}^{2 \pi} \  e^{i x \cos \phi }
  \ d \phi \ \ \ \ \
 J_{1}(x) \ =  -\frac{1}{2 \pi}
   \ \int_{0}^{2 \pi} \  e^{i x \cos \phi } \  \sin{\phi}
  \ d\phi\,,
 \label{J0}
 \end{eqnarray}
 the representation of spin-non-flip and spin flip amplitude is
 \begin{eqnarray}
 \Phi^{h}_{1}(s,t) \ = \ i
   \ \int_{0}^{\infty} \ b J_{0}(b q) \frac{\chi_{0}(s,b)}{1+\chi_{0}(s,b)} db \,;
 \label{K-matrix}
 \end{eqnarray}
 \begin{eqnarray}
 \Phi^{h}_{5}(s,t) \ = \ i
   \ \int_{0}^{\infty} \ b^2 J_{1}(b q) \frac{\chi_{sf}(s,b)}{\left(1+\chi_{0}(s,b\right))^2} db \,.
 \label{K-m-spin}
 \end{eqnarray}

 \section{\boldmath $U_T$-matrix unitarisation}
 If Eq.~(\ref{log-eq}) has  an additional coefficient $n$
 \begin{eqnarray}
 \frac{dN}{dy} =\Delta N \ [ \ 1 - N/n \ ]\,. \label{UT-eq}
 \end{eqnarray}
 we obtain, for $n=2$, the unitarisation in
 the standard $U$-matrix  form intensively  explored in~\cite{Chrustalev}.

 In the impact parameter representation, the properties of the $U$-matrix are  explored in
  \cite{TT1}. In this scheme, the hadronic amplitude is given by
  \begin{eqnarray}
 \Phi^{h}(s,t) \ = \ \frac{i}{2  \pi}
   \ \int \ d^2b  \  e^{i \vec{b} \vec{q} } \frac{\chi(s,b)}{1+ \chi(s,b)/2}\,,
 \label{U-mat}
 \end{eqnarray}
 where $\chi(s,b)$ is the same Born amplitude as before.

 Comparing Eq.~(\ref{U-mat}) with Eq.~(\ref{K-mat}), we see that both have
 the same rational form but differ by the additional coefficient in the denominator.
 This additional coefficient leads to different analytic properties: the upper bound
 at which the overlapping function saturates
 will be in twice as large as in the eikonal or $U_e$ representations,
 and the inelastic
 overlap function at $b=0$ will be tend to zero at high energies. This leads
 to the new relation \cite{TT1} $\sigma_{\rm el}/\sigma_{\rm tot} \rightarrow 1$.

 For the helicity amplitudes of $pp$ scattering,
 the corresponding solution of the unitarity equations \cite{TT-spin}:
 \begin{eqnarray}
 \Phi_{\lambda_3,\lambda_4,\lambda_1,\lambda_2}({\bf p},{\bf q}) & =
  & U_{\lambda_3,\lambda_4,\lambda_1,\lambda_2}({\bf p},{\bf q})+ \label{heq}\\
 & & i\frac{\pi}{8}
 \sum_{\lambda ',\lambda ''}\int d\Omega_{{ \bf \hat k}}
 U_{\lambda_3,\lambda_4,\lambda ',\lambda ''}({\bf p},{\bf k})
 \Phi_{\lambda ',\lambda '',\lambda_1,\lambda_2}({\bf k},{\bf q})\,,\nonumber
 \end{eqnarray}
 In the impact parameter representation, one obtains the following equations
 relating the unitarised helicity amplitudes $\Phi_i(s,b)$ to the Born
 amplitudes $u_i(s,b)$ \cite{TT-spin}:
 \begin{eqnarray}
 \Phi_1 & = & \frac{(u_1 + u_1^2 - u_2^2)(1 + u_3 + u_4) - 2(1 + 2u_1 - 2u_2)u_5^2}
 {(1 + u_1 - u_2)\left[(1 + u_1 + u_2)(1 + u_3 + u_4)- 4u_5^2\right]}\,,\nonumber\\
 \Phi_2 & = & \frac{u_2(1 + u_3 + u_4) - 2u_5^2}
 {(1 + u_1 - u_2)\left[(1 + u_1 + u_2)(1 + u_3 + u_4)- 4u_5^2\right]}\,,\nonumber\\
 \Phi_3 & = & \frac{(u_3 + u_3^2 - u_4^2)(1 + u_1 + u_2) - 2(1 + 2u_3 - 2u_4)u_5^2}
 {(1 + u_3 - u_4)\left[(1 + u_1 + u_2)(1 + u_3 + u_4)- 4u_5^2\right]}\,,\nonumber\\
 \Phi_4 & = & \frac{u_4(1 + u_1 + u_2) - 2u_5^2}
 {(1 + u_3 - u_4)\left[(1 + u_1 + u_2)(1 + u_3 + u_4)- 4u_5^2\right]}\,,\nonumber\\
 \Phi_5 & = & \frac{u_5}
 {(1 + u_1 + u_2)(1 + u_3 + u_4)- 4u_5^2}\,,\label{fi}
 \end{eqnarray}
 where for simplicity we omitted the arguments in the functions $\Phi_i(s,b)$
 and $u_i(s,b)$.
 If we take $\chi_{c} = u_1+ u_3$ and $\chi_{sf}=u_5$
  in the same approximation as in the $U_e$ case, the spin-non-flip and
 spin-flip amplitude will be
 \begin{eqnarray}
 \Phi^{h}_{1}(s,t) \ = \ i
   \ \int_{0}^\infty \ b J_{0}(b q) \frac{\chi_{c}(s,b)}{1+\chi_{c}(s,b)/2} db \,;
 \label{tot03}
 \end{eqnarray}
 \begin{eqnarray}
 \Phi^{h}_{5}(s,t) \ = \  i
   \ \int_{0}^{\infty} \ b^2 J_{1}(b q) \frac{\chi_{sf}(s,b)}{\left(1+\chi_{c}(s,b)/2\right)^2} db \, .
 \label{tot04}
 \end{eqnarray}
  It is clear that these forms can also be obtained by the same procedure as we used
 in the case   of $U_e$ unitarisation.

 \section{Eikonal unitarisation scheme}
 To obtain the standard eikonal representation of the elastic scattering amplitude
 in the impact parameter representation one must take the non-linear equation
 in the form \cite {dif04}
 \begin{eqnarray}
 \frac{dN_e}{dy} = -\Delta \log(1-N_e) [1-N_e]\,. \label{nl-eik}
 \end{eqnarray}
 where $y =\log(s/s_0)$ and where the subscript ``e" implies that the solution $N_e$
 has exactly the standard eikonal form as shown in ref. \cite{dif04}, {\it i.e.}
 \begin{eqnarray}
 N_e = \Gamma (s,b)= [1- e^{- \chi(s,b)}]. \label{nl-eik1b}
 \end{eqnarray}

  The eikonal representation is then
  \begin{eqnarray}
 \Phi^{h}(s,t) \ =  \frac{i}{2  \pi}
   \ \int \  e^{i \vec{b} \vec{q} } \ \left[1 - e^{- \chi(s,b) }\right]
  \ d^2 b,
 \label{tot0}
 \end{eqnarray}
 where the eikonal phase in the case of spin-dependent potential has a matrix structure
 and the quasi-potential $V(s,r)$ contains the non-spin-flip part and , for example,
 spin-orbital and spin-spin interaction:
 \begin{eqnarray}
  \chi(s,b) = \chi_{0}(s,b)
 - i \ \vec{n} \cdot (\vec{\sigma}_{1} + \vec{\sigma}_{2} ) \chi_{LS}(s,b)
- i  (\vec{\sigma}_{1} \cdot \vec{\sigma}_{2} ) \ \chi_{SS}(s,b).
\end{eqnarray}

 Taking into account the Eqs. (\ref{J0}) and (\ref{J1}), we have for the spin non-flip
 \begin{eqnarray}
 \Phi^{h}_{1}(s,t) \ =  \  i
   \ \int_{0}^{\infty} \ b J_{0}(b q)\left[ 1- e^{\chi_{0}(s,b)}\right] \
   \left[1 - b^2 \chi^2_{\rm LS}(s,b) -3/2 \chi^2_{\rm SS}(s,b)\right]
  \ d b,
 \label{J1}
 \end{eqnarray}
 and for the spin-flip
 \begin{eqnarray}
 \Phi^{h}_{5}(s,t) \ =   \ i
   \ \int_{0}^{\infty} \  J_{1}(b q) \chi_{1} e^{\chi_{0}(s,b)}
 \ b \  \left[\chi_{\rm LS}(s,b) +i \ \chi_{\rm LS}(s,b) \ \chi_{\rm SS}(s,b)\right]  \ d b,
 \label{tot0a}
 \end{eqnarray}
 where
 \begin{eqnarray}
 \chi(s,b)_{0}  &\approx &
   \ \int_{-\infty}^{\infty}  V_{0}(s,b,z)   dz\,; \\
 \chi(s,b)_{1}  &\approx  & \frac{b}{2}
   \ \int_{-\infty}^{\infty}  V_{1}(s,b,z)   dz\,.
 \label{potential5}
 \end{eqnarray}
  If, for example,
   the potentials $V_{0}$ and $V_{1}$ are assumed to be Gaussian
  \begin{eqnarray}
 V(s,b)_{0,1} \ \approx
   \ \int_{-\infty}^{\infty} e^{-B r^2}  \ dz = \frac{\sqrt{\pi}}{\sqrt{B}} e^{-B b^2}\,,
 \label{potential2}
 \end{eqnarray}
 in the first Born approximation, $\Phi_{0}^{h}$ and $\Phi_{1}^{h}$
 will have the forms
 \begin{eqnarray}
 \Phi^{h}_{1}(s,t) \ \approx
   \ \int_{0}^{\infty} \ b J_{0}(b q) e^{-B b^2} d b = e^{-B q^2};
 \label{tot0b}
 \end{eqnarray}
 \begin{eqnarray}
 \Phi^{h}_{5}(s,t) \ \approx
   \ \int_{0}^{\infty} \ b^2 J_{1}(b q) e^{-B b^2} d b =  q  B e^{-B q^2}.
 \label{tot01}
 \end{eqnarray}

 \section{The analysing power in the different unitarisation schemes}
   Now let us compare the spin correlation parameter $A_N$ in the different
   unitarisation schemes.
   For that we use the Born terms of the spin-non-flip and spin-flip
   Born terms of the proton-proton elastic scattering calculated in the
   framework of the Dubna Dynamical Model (DDM) \cite{selmog1,DDM}.
%
 \vspace{-1.5cm}
 \begin{figure}[!h]
 \begin{flushleft}
 \epsfysize=55mm
 \epsfbox{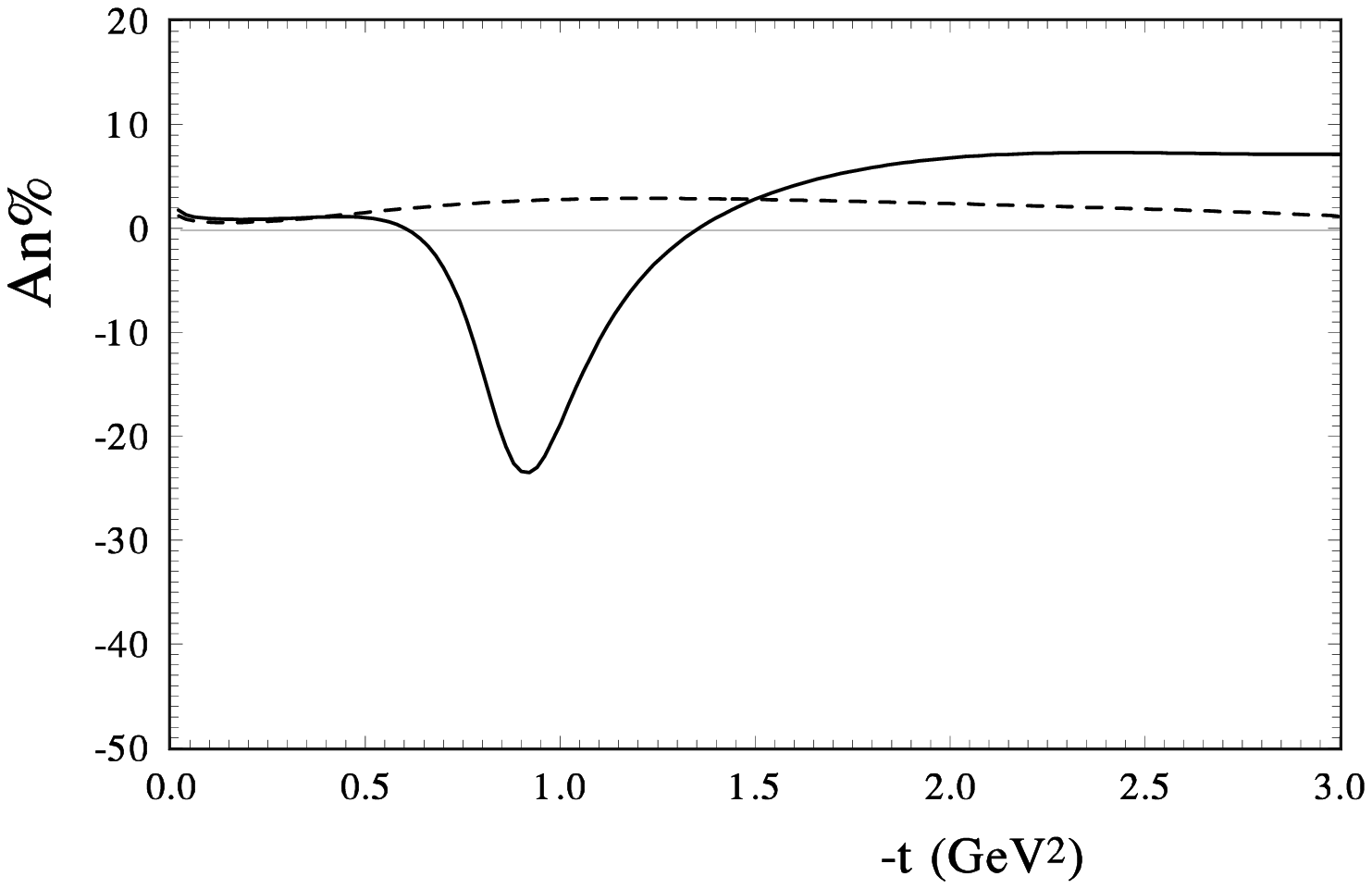}
 \end{flushleft}
 \vspace{-6.0cm}
 \begin{flushright}
 \epsfysize=55mm
 \epsfbox{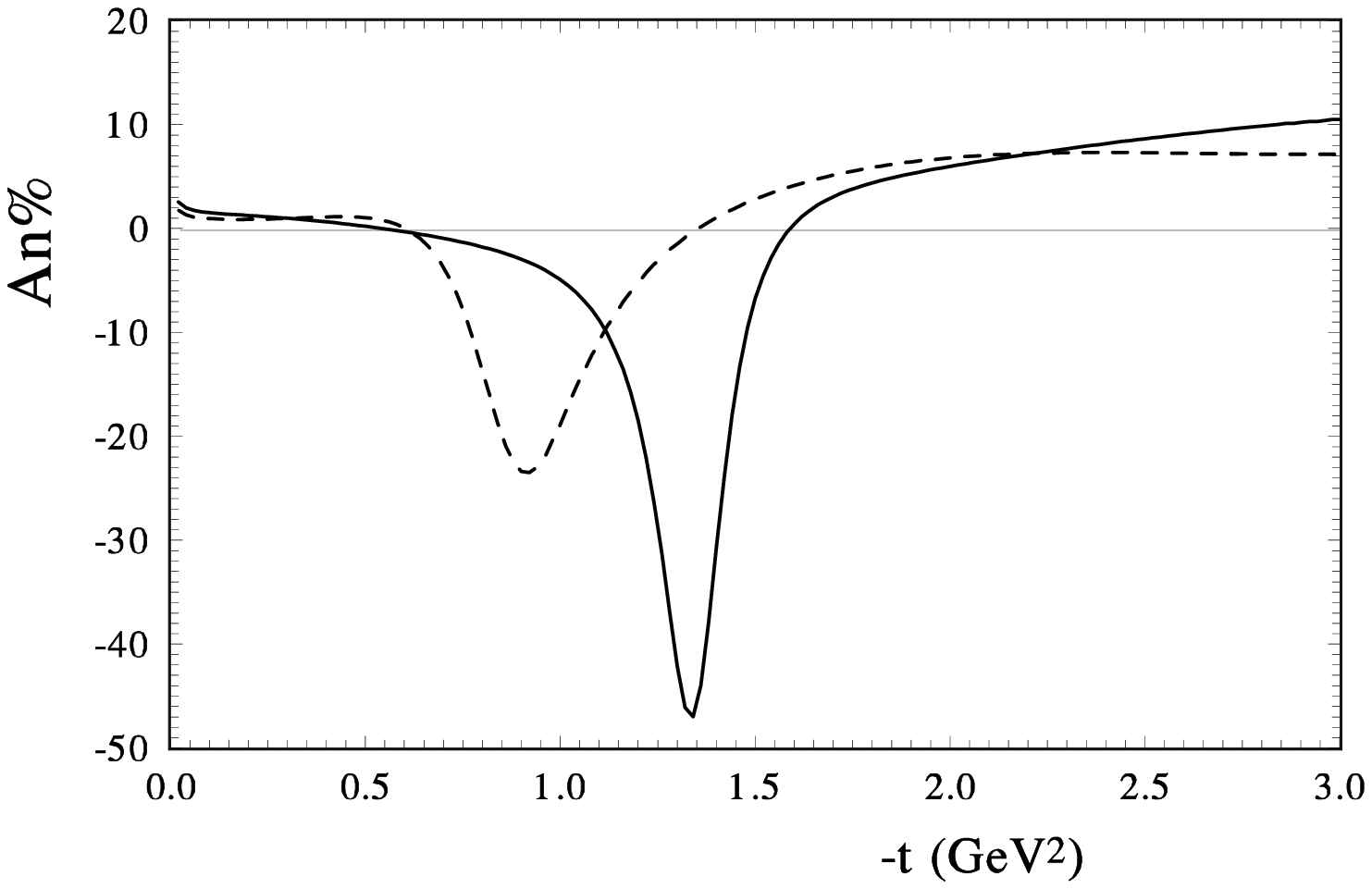}
 \end{flushright} \vspace{-3mm}
 \caption{{\bf a}[left]  $A_N(s,t)$ for
 the Born amplitudes calculated in the DDM \cite{DDM-spin}
  at $\sqrt{s}=50\,$GeV (full line) and
 at $\sqrt{s}=500\,$GeV (dashed line);
 {\bf b}[right]  $A_N(s,t)$ in the case of the  eikonal unitarisation
 with the Born amplitudes calculated in the DDM \cite{DDM-spin}
  at $\sqrt{s}=50\,$GeV (full line) and
 at $\sqrt{s}=500\,$GeV (dashed line).
 } \vspace{-5mm}
 \end{figure}
 This model, which takes into account the interactions at large distances, predicts
 non-vanishing spin effects at high energies \cite{DDM-spin}.
 The values of $A_N(s,t)$ corresponding to the  Born terms
 are shown in Fig.~1a.
 We can see that whereas the analysing power is not small at $\sqrt{s}=50\,$GeV,
 it is negligible at $\sqrt{s}=500\,$GeV.
 Now we can use the eikonal form of the  unitarisation.
 Our results are shown in Fig.~1b. We see that in this case
 the size of  $A_N$ grows and now it is a~measurable effect
 up to $\sqrt{s}=500\,$GeV. This result is linked to the fact
 that the diffractive structure of proton-proton scattering does
 not  disappear at this energy.

 Now we can consider the unitarisation procedure in the form of the  $U_e$-matrix,
 Eqs.~(\ref{K-matrix}, \ref{K-m-spin}).
 The result of the calculation is shown in Fig.~2. The size of $A_N(s,t)$ is
 above that in the case of the eikonal unitarisation. It remains
 large at $\sqrt{s}= 500\,$GeV. The size of $A_N$ is positive and large
 after the diffraction minimum, and reaches 20\,\% at $|t|= 2\,$GeV$^2$.
 However, a~comparison with the eikonal unitarisation (see Fig.~1b)
 shows that we have practically the same form of the analysing power
 in both cases. So, the difference is only quantitative but not qualitative. Despite an
 essential change in the functional form of the  unitarisation procedure, we obtain very
 similar results.

 \vspace{-1.5cm}
 \begin{figure}[!h]
 \epsfysize=60mm
 \centerline{\epsfbox{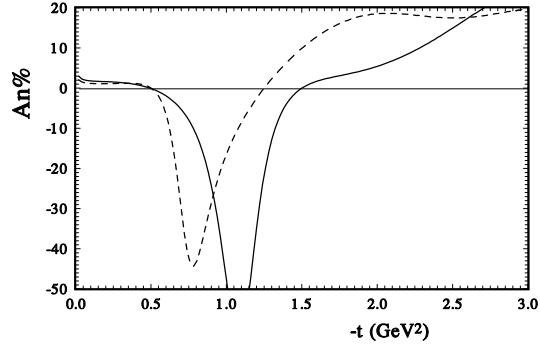}}
 \caption{ $A_N(s,t)$ in the case of the  $U_e$-matrix unitarisation
 with the Born amplitudes calculated in the DDM \cite{DDM-spin}
  at $\sqrt{s}=50\,$GeV (full line) and
 at $\sqrt{s}=500\,$GeV (dashed line).}
 \end{figure} \vspace{-3mm}

 A very different picture, presented in Fig.~3a, is obtained if we use
 $U_T$-matrix unitarisation. In this case, $A_N(s,t)$ has a~very different
 form, coming mainly from the form of the spin-non-flip amplitude,
 which for the Born term of the DDM does not reproduce elastic proton-proton scattering.

 \vspace{-1.5cm}
 \begin{figure}[!ht]
 \begin{center}
 \epsfysize=48mm
 \epsfbox{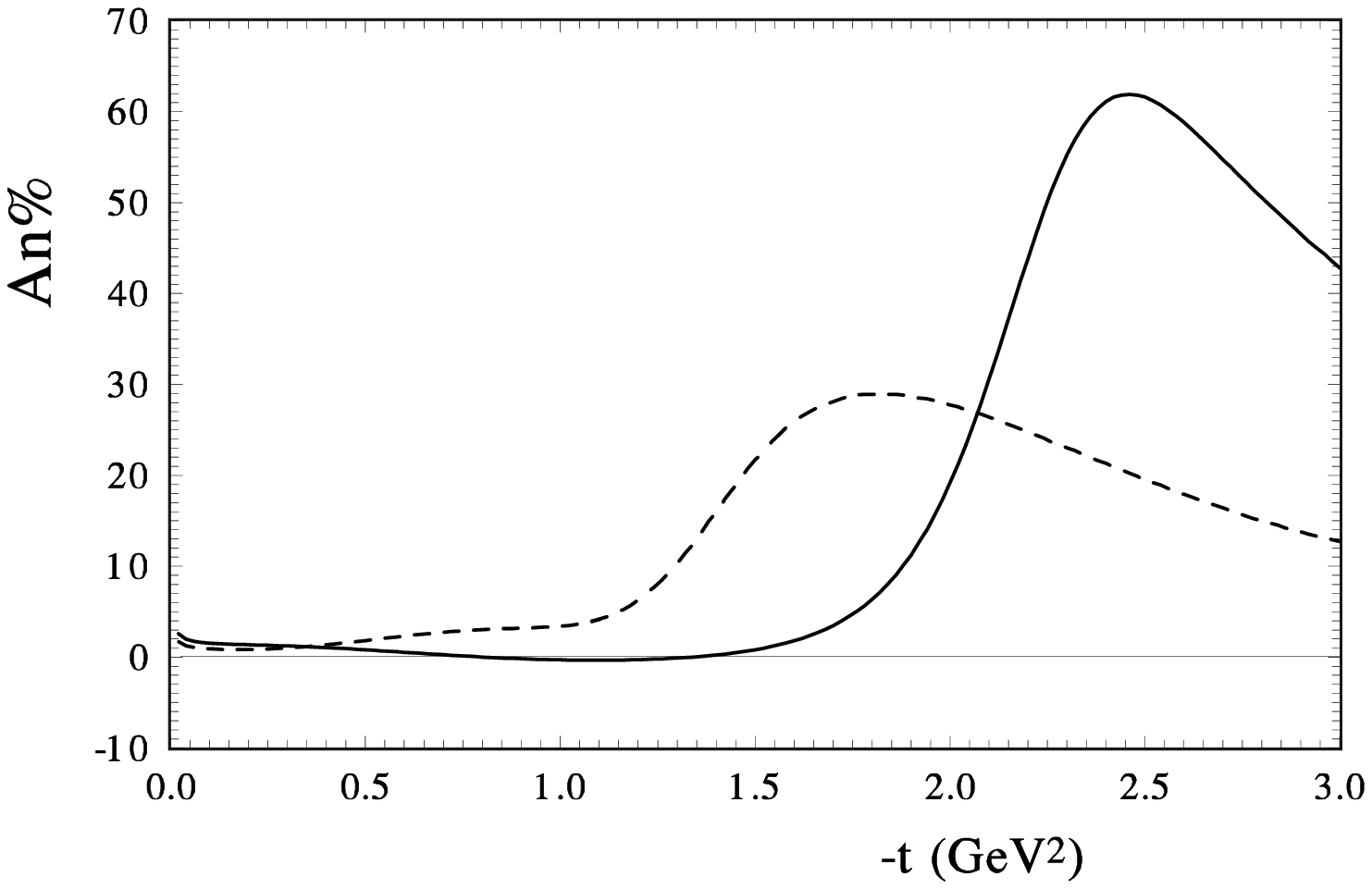}\ \ \
 \epsfysize=48mm
 \epsfbox{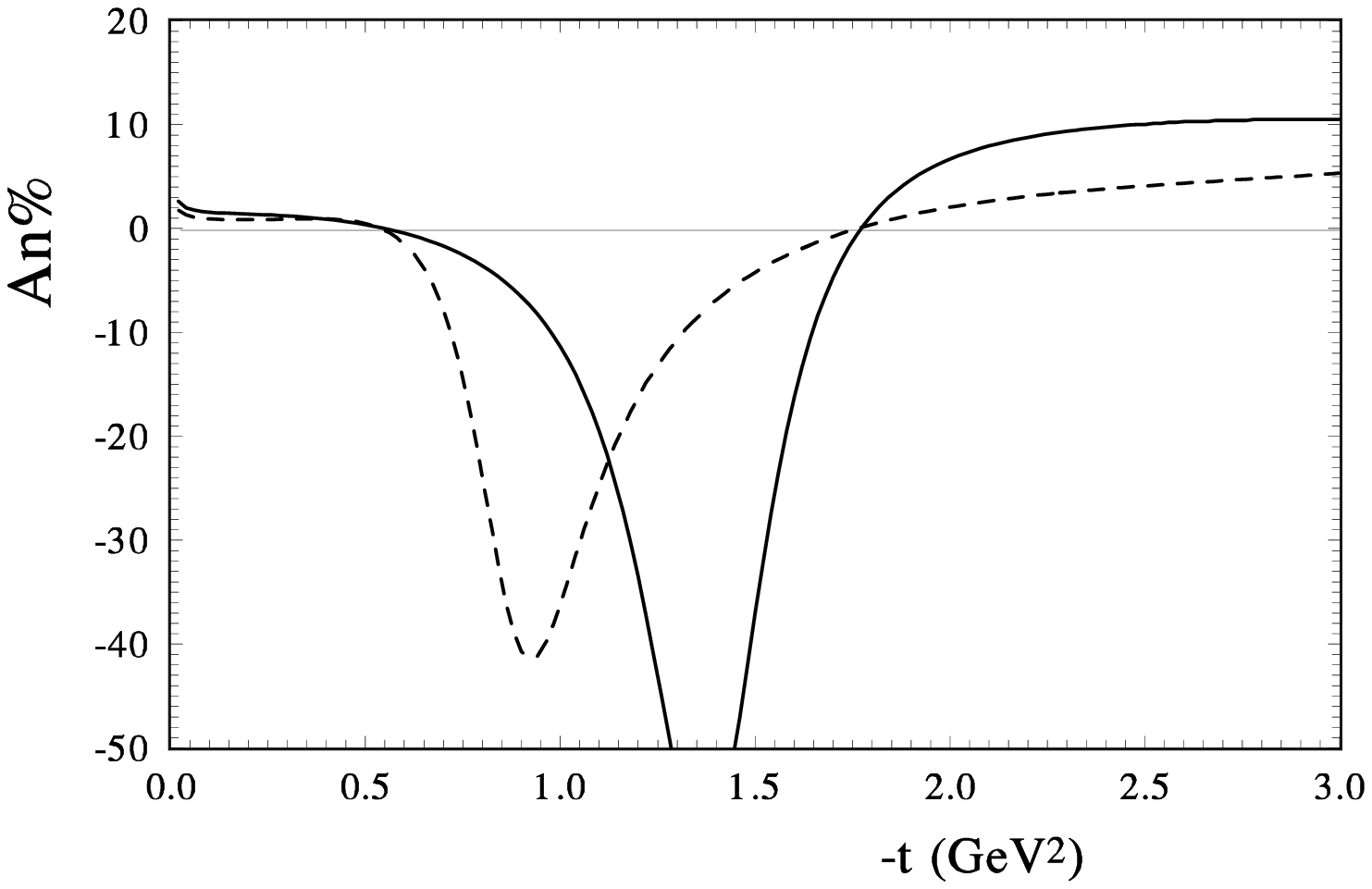}
 \end{center}
 \caption{{\bf a}[left]
 $A_N(s,t)$ in the case of the  $U_T$-matrix unitarisation
 with the Born amplitudes calculated in the DDM \cite{DDM-spin}
  at $\sqrt{s}=50\,$GeV (full line) and
 at $\sqrt{s}=500\,$GeV (dashed line);
 {\bf b})[left] $A_{N}(s,t)$ in the case of the  $U_T$-matrix unitarisation
 with the new Born amplitudes for $U_T$  calculated by new fit
  at $\sqrt{s}=50\, $GeV (full line) and
 at $\sqrt{s}=500\,$GeV (dashed line)}
 \end{figure} \vspace{-.5cm}

 In order to fix this problem, we made a new fit of the scattering amplitude
to obtain the correct  description the differential cross sections at high
energy in the framework of the  $U_T$-matrix unitarisation.  The new Born
leads to the analysing power shown in Fig.~3b. The resulting Born term is
rather different from that in the eikonal or $U_e$ cases,
but we see that all the unitarisation schemes
lead to similar curves for the analysing power.

    \section{Conclusion}
From the above analysis, we are led to the conclusion that the unusual properties
 of the $U_T$ matrix unitarisation are not connected with its functional form.
Other rational forms, such as the $U_e$ matrix, have the same properties as
the eikonal.

 The spin-non-flip and spin-flip amplitudes also have very similar functional
 forms in the $U_T$ and $U_e$
 schemes, differing only by an additional coefficient in the denominator.
  The comparison  of the polarisation effects calculated for different unitarisation
  shows that the eikonal and $U_e$ matrix qualitatively give the same results for the
  same Born term, but that a very different Born term needs to be used in the
 $U_T$-matrix case, suggesting that the underlying dynamical picture of the scattering must be quite different.

\section*{Acknowledgements}
{\small The authors would like to thank  for helpful discussions
  J. Fischer and E. Predazzi.
 O.S. gratefully acknowledges financial support
  from FRNS and would like to thank the  University of Li\`{e}ge
  where part of this work was done.
    }


\begin{thebibliography}{99}
\bibitem{mog1}V. Barone, E. Predazzi, in the book ``High Energy particle Diffraction", Springer,
2002, ISBN 3540421076;
%
C.~Bourrely, J.~Soffer and T.\,T.~Wu,
  Phys.\ Rev.\  D {\bf 19} (1979) 3249;
%
 A.\,F.~Martini and E.~Predazzi,
  Phys.\ Rev.\  D {\bf 66} (2002) 034029
  [arXiv:hep-ph/0209027]
%
\bibitem{selmog1}
  S.\,V.~Goloskokov, S.\,P.~Kuleshov and O.\,V.~Selyugin,
  Z.\ Phys.\  C {\bf 50} (1991) 455
%
\bibitem{dif04}
J.\,R.~Cudell and O.\,V.~Selyugin,
Czech.\ J.\ Phys.\  {\bf 54} (2004) A441
[arXiv:hep-ph/0309194]
%
\bibitem{Predazzi66} E. Predazzi, Ann. of Phys., {\bf 36} (1966)
228, 250
%
\bibitem{mog2}
B.\,Z.~Kopeliovich and B.\,G.~Zakharov,
Phys.\ Lett.\  B {\bf 226}  (1989) 156;
%
M.~Anselmino and S.~Forte,
Phys.\ Rev.\ Lett.\  {\bf 71} (1993) 223
[arXiv:hep-ph/9211221];
%
  S.\,V.~Goloskokov,
  Phys.\ Lett.\  B {\bf 315} (1993) 459;
%
  A.\,E.~Dorokhov, N.\,I.~Kochelev and Yu.\,A.~Zubov,
  Int.\ J.\ Mod.\ Phys.\  A {\bf 8} (1993) 603;
%
  J.\,R.~Cudell, E.~Predazzi and O.\,V.~Selyugin,
  Phys.\ Part.\ Nucl.\  {\bf 35} (2004) S75
  [arXiv:hep-ph/0312195];
  Eur.\ Phys.\ J.\  A {\bf 21} (2004) 479
  [arXiv:hep-ph/0401040]
%
\bibitem{DDM-spin}
  N.~Akchurin, S.\,V.~Goloskokov and O.\,V.~Selyugin,
  Int.\ J.\ Mod.\ Phys.\  A {\bf 14} (1999) 253
%
\bibitem{Shiff} L.\,I. Shiff, Phys.Rev. {\bf 103} (1956) 443
%
\bibitem{Goldberger} M.\,L. Goldberger and K.\,M. Watson, in the book ``Collision Theory",
Dover Publications, 2004, ISBN: 0486435075
%
\bibitem{Blokhintzev} D.\,I.~Blochinzev, Nuov. Cim. {\bf 30} (1968) 1024
%
\bibitem{Barbashev} B.\,M.~Barbashov, D.\,I.~Blochinzev, V.\,V. Nesterenko,
 V.\,I. Pervushin, Phys. Part.  Nucl. {\bf 4} (1973) 623
%
\bibitem{Martiros}
  K.\,A.~Ter-Martirosyan,
  Pisma Zh.\ Eksp.\ Teor.\ Fiz.\  {\bf 15} (1972) 734
%
\bibitem{Kaidalov1}   A.\,B.~Kaidalov, L.\,A.~Ponomarev and K.\,A.~Ter-Martirosian,
  Yad.\ Fiz.\  {\bf 44} (1986) 722
  [Sov.\ J.\ Nucl.\ Phys.\  {\bf 44} (1986) 468]
%
\bibitem{Martynov1}  M.~Giffon, E.~Martynov and E.~Predazzi,
  Z.\ Phys.\  C {\bf 76} (1997) 155
%
\bibitem{Capella}
  A.~Capella, A.\,P.~Contogouris and J.~Tran Thanh Van,
  Phys.\ Rev.\  {\bf 175} (1968) 1892
%
\bibitem{Lehar}
  J.~Bystricky, F.~Lehar and P.~Winternitz,
  J.\ Phys.\ (France) {\bf 39} (1978) 1;
   C.~Lechanoine-LeLuc and F.~Lehar,
  Rev.\ Mod.\ Phys.\  {\bf 65} (1993) 47
%
\bibitem{WY}
  G.\,B.~West and D.\,R.~Yennie,
  Phys.\ Rev.\  {\bf 172} (1968) 1413;
  R.~Cahn,
  Z.\ Phys.\  C {\bf 15} (1982) 253
%
 \bibitem{Selphase}
   O.\,V.~Selyugin,
  Phys.\ Rev.\  D {\bf 60} (1999) 074028
%
\bibitem{Morel}
G.~Cohen-Tannoudji, Ph. Salin, and A. Morel,
 Nuov. Cim.  {\bf 55 A} (1968) 412
%
\bibitem{mog3}
  L.\,D.~McLerran and R.~Venugopalan,
  Phys.\ Rev.\  D {\bf 50} (1994) 2225
  [arXiv:hep-ph/9402335];
  I.~Balitsky,
  Nucl.\ Phys.\  B {\bf 463} (1996) 99
  [arXiv:hep-ph/9509348];
%
  Y.\,V.~Kovchegov,
  Phys.\ Rev.\  D {\bf 60} (1999) 034008
  [arXiv:hep-ph/9901281];
%
  J.~Jalilian-Marian, A.~Kovner, A.~Leonidov and H.~Weigert,
  Phys.\ Rev.\  D {\bf 59} (1999) 014014
  [arXiv:hep-ph/9706377]
%
\bibitem{d-d-m}
  A.\,H.~Mueller,
  Nucl.\ Phys.\  B {\bf 437} (1995) 107
  [arXiv:hep-ph/9408245]
%
\bibitem{CPS-Prag}
  O.\,V.~Selyugin, J.\,R.~Cudell and E.~Predazzi,
  Eur.\ Phys.\ J.\ ST {\bf 162} (2008) 37
  [arXiv:0712.0621 [hep-ph]]
%
 \bibitem{Chrustalev} V.\,I. Savrin, N.\,E.~Tyurin,  O.\,A. Chrustalev:
 Fiz. Elem. Chast. At. Yadra  {\bf 7} (1976) 21
%
 \bibitem{TT1}
   S.\,M.~Troshin and N.\,E.~Tyurin,
  Phys.\ Lett.\  B {\bf 316} (1993) 175
  [arXiv:hep-ph/9307250]
%
 \bibitem{TT-spin}   S.\,M.~Troshin and N.\,E.~Tyurin,
  in the Procedings of the 16th International Spin Physics Symposium (SPIN 2004),
 Trieste, Italy,  Oct 10--16,  2004, *Trieste/Mainz 2004,
 SPIN 2004*, 523--526, arXiv:hep-ph/0412207
%
 \bibitem{DDM}   S.\,V.~Goloskokov, S.\,P.~Kuleshov and O.\,V.~Selyugin,
  Fiz.\ Elem.\ Chast.\ Atom.\ Yadra {\bf 18} (1987) 39;
    Sov.\ J.\ Part.\ Nucl.\  {\bf 18} (1987) 14
%
 \bibitem{CS-sat06}
  O.\,V.~Selyugin and J.\,R.~Cudell,
  Czech.\ J.\ Phys.\  {\bf 56} (2006) F237
  [arXiv:hep-ph/0611305]
%
 \bibitem{CS-BDL06}
  J.\,R.~Cudell and O.\,V.~Selyugin,
  Phys.\ Lett.\  B {\bf 662} (2008) 417
  [arXiv:hep-ph/0612046]
%
\end{thebibliography}
\end{document}